\documentstyle[twoside,fleqn,npb,epsfig]{article}
\def\Journal#1#2#3#4{{#1} {\bf #2}, #3 (#4)}

\def\PRD{{\em Phys. Rev.} D}

\def\be{\begin{equation}}
\def\ee{\end{equation}}
\def\bea{\begin{eqnarray}}
\def\eea{\end{eqnarray}}

\def\MW{m_{\rm W}}
\def\Mt{m_{\rm t}}
\def\MH{m_{\rm H}}

\def\MA{m_{\rm A}}
\def\GH{\Gamma_{\rm H}}
\def\GW{\Gamma_{\rm W}}

\def\pt{p_{\rm T}}
\def\Et{E_{\rm T}}
\def\sineff{\sin^2\!\theta_{\rm eff}^{\rm {lept}}}

\newcommand{\AmS}{{\protect\the\textfont2
  A\kern-.1667em\lower.5ex\hbox{M}\kern-.125emS}}

\title{Precision Physics at the LHC}
\author{Fabiola~Gianotti\address{CERN, EP Division, 1211 Gen\`eve 23, \\  
 Gen\`eve, Switzerland \\E-mail: Fabiola.Gianotti@cern.ch} \\
        and
        Monica~Pepe~Altarelli\address{Laboratori Nazionali dell'INFN (LNF-INFN) \\
        00044 Frascati, Italy \\E-mail:
        Monica.Pepe.Altarelli@cern.ch}}
\begin{document}
\begin{abstract}

\begin{center}
{\normalsize Talk given by Monica Pepe Altarelli at the \\
5th Zeuthen Workshop on Elementary Particle Theory on}\\
{\large ``Loops and Legs in Quantum Field Theory''}\\
{\normalsize Bastei/K\"{o}nigstein, Germany, April 2000}\\ 
\end{center}

A large number of precision measurements
  will be possible with the ATLAS and CMS experiments 
  at the CERN Large Hadron Collider (LHC). 
  Examples from W  physics, Drell-Yan production of lepton pairs, Triple-Gauge Couplings,
  top physics, Higgs and Supersymmetry are discussed.
\end{abstract}

\maketitle

    It is well known since many years that the LHC has a large
   discovery potential for new physics, e.g. Higgs and Supersymmetry (SUSY), 
    owing to the large centre-of-mass energy ($\sqrt{s}$=14 TeV) and
   high design luminosity ($10^{34}$ cm$^{-2}$ s$^{-1}$). 
   
    In addition, the two general-purpose proton-proton experiments
   ATLAS~\cite{ATLAS} and CMS~\cite{CMS} will be able to
   perform precision measurements 
   in a large number of physics 
   channels~\footnote{LHCb, the dedicated B-physics
   experiment, is not discussed in this paper.}. In most
   cases, these measurements 
   are expected to improve significantly on the 
   results obtained  at previous machines (LEP and TeVatron).
   
     Precision physics with ATLAS and CMS is the subject of this paper.
     After a discussion of the most relevant issues for
    precision physics at LHC (Sect.~\ref{key}), 
     examples of measurements for some physics channels are presented:
    W-mass (Sect.~\ref{wmass}), Drell-Yan production of lepton pairs (Sect.~\ref{Drell-Yan}),
    Triple-Gauge Couplings (Sect.~\ref{TGC}),
    top physics (Sect.~\ref{top}), Higgs (Sect.~\ref{Higgs}) and
    SUSY (Sect.~\ref{SUSY}) (see Refs.~\cite{TDR,LHCEW,LHCtop}
    for extensive and recent reviews of these subjects).
                            
    The following assumptions on the instantaneous and
   integrated luminosities are made throughout this paper. The initial
   luminosity is expected to be $10^{33}$~cm$^{-2}$~s$^{-1}$ 
   (hereafter called ``low luminosity") and
   should rise, during the first 
   three years of operation, to the design luminosity
   of 10$^{34}$ cm$^{-2}$ s$^{-1}$ (hereafter called ``high luminosity").
    Integrated luminosities of 10~fb$^{-1}$, 30~fb$^{-1}$, 100~fb$^{-1}$
   and 300~fb$^{-1}$ should be collected after
   one year, three years, four years and less than ten years of data taking, 
   respectively. 
  

\section{Key issues\label{key}}
  
  The main asset for precision physics at the LHC is statistics. 
  
  Table~\ref{tab:rates} shows the expected rates of some
 representative physics processes, both from Standard Model (SM)
 and new physics.   
\begin{table}[bt]
\caption{For the physics channels listed in the first column, 
 the cross section and 
 the approximate expected number of events in each experiment 
in one second and 
 in one year at low luminosity ($10^{33}$~cm$^{-2}$~s$^{-1}$). \label{tab:rates}}
\begin{center}
\footnotesize
\begin{tabular}{lccc}
\hline
         &               &               & \\[-.4pc]
Process  & $\sigma$ (pb) & Events/second & Events/year  \\
         &               &               & \\[-.4pc]
\hline
         &               &               & \\[-.4pc]
$\rm W\rightarrow e\nu$    & $1.5\times 10^4$  & 15       & 10$^{8}$ \\
$\rm Z\rightarrow e^+e^- $ & $1.5\times 10^3$  & 1.5      & 10$^{7}$ \\
$\rm{t\overline{t}}$        &  800              & 0.8      & 10$^{7}$ \\
$\rm{b\overline{b}}$        &  5$\times 10^{8}$ & $5\times 10^{5}$ & $\ 10^{12}$ \\
$\rm \tilde{g}\tilde{g}$ (m$_{\tilde g}$=1 TeV) & 1  & 10$^{-3}$  & 10$^{4}$ \\
H ($\MH$=700 GeV)    & 1                 & 10$^{-3}$ & 10$^{4}$  \\
Inclusive jets         &        &   &    \\[-.5pc]  
                       & 10$^5$         & 10$^{2}$  & 10$^{9}$ \\ [-.4pc]
$\pt >$~200 GeV        &                &           &          \\
\hline
\end{tabular}
\end{center}
\end{table}
   In the initial phase at low luminosity,  almost 50
   W and five Z bosons decaying to lepton pairs will be 
   produced every second, as well as one $\rm{t\overline{t}}$
   pair and 500,000 $\rm{b\overline{b}}$ pairs. One SM Higgs
   boson of 700~GeV mass and one pair of 1~TeV
   gluinos  would be produced every 15 minutes,
   while the rate of QCD jets with $\pt>$~200~GeV will be
   about 100 Hz. This last process is
    expected to be one of the dominant
    backgrounds to many interesting physics channels.  
   Integrated over one year of data taking at low luminosity, these
   rates give rise to samples of millions of events in almost all
   channels. The LHC can therefore be considered as a factory
   of a large number of particles: W  and Z bosons, 
   top and b quarks, and possibly also Higgs boson(s) and 
   supersymmetric particles.

    As a consequence, for most measurements 
 performed at the LHC the statistical error and the component
  of the systematic error which scales
   as $1/\sqrt{N}$ will be negligible (where $N$ is the number
  of selected events). The 
  uncertainty will instead be dominated by the component of the systematic
  error, arising from the knowledge of both detector and physics,
  which depends only weakly on the number of events.    
   However, large statistics will allow 
  hard cuts to be applied in order to select clean and well-understood events. 
   Furthermore, high-statistics ``control samples",
  e.g. Z~$\rightarrow\ell\ell$ decays, will be available to study
  the detector response and the physics 
  (background shapes, $\pt$ distributions, etc.) in great detail. 
%
%
                                  
     Three main sources of uncertainty are expected to affect
 precision measurements at the LHC:
\begin{itemize}
\item The lepton energy and momentum scale, 
that is related to the calibration of the inner detector,
    of the electromagnetic calorimeter and of the
   muon spectrometer. This is the dominant source of
   uncertainty on the W  mass measurement
   at the TeVatron, where the absolute lepton scale
   is known with a precision of $\sim$0.1\%~\cite{CDF,D0}. 
    At the LHC such a precision will be
   adequate for most measurements except for
    the W mass, for which a much better
   knowledge, i.e. $\sim$0.02\%, will be needed in order to improve on the 
   accuracy expected at the end of the
    LEP2 and TeVatron programs, as described in Sect.~\ref{wmass}. 
    The lepton scale will be determined {\it in situ} 
     by using, for instance, 
   the large-statistics samples of Z~$\rightarrow\ell\ell$ decays.
    The Z boson has the advantage of being a resonance very 
    close in mass
   to the W  and to the h boson of the Minimal Supersymmetric
   Standard Model (MSSM), so that the
   extrapolation error from the calibration region
   to the measurement region 
   is considerably reduced. In contrast, 
    TeVatron experiments do not have today 
    enough Z events to calibrate the lepton scale with high accuracy,
    and other resonances like ${\rm J/\psi}$ or $\pi^0$ have  to be used. 
    Preliminary studies performed in ATLAS~\cite{TDR}
    indicate that a precision
    of 0.02\% will be very difficult to achieve but not impossible.  
     
\item The jet energy scale, contributing
  for instance to the uncertainty on the top mass. Unlike  the
  lepton scale, the precision on the jet absolute scale depends not only on 
  the detector (calorimeter in this case) calibration,
  but also on the knowledge of the physics (fragmentation, gluon 
  radiation, etc.). Today at the TeVatron the jet scale is determined 
  with a precision of about 3\%~\cite{D0j}, 
  by using mainly events with a  $\gamma$ or a Z decaying into leptons
  balanced by one high $\pt$ jet.
  The LHC goal is to reach about 1\%.
   In addition to the TeVatron methods, at the LHC
  the light quark jet calibration will be based on
  W$\rightarrow$jj decays from $\rm{t\rightarrow b}$W. Indeed,
   $\rm{t\overline{t}}$ final states where one top decays
   to b$\ell\nu$ and the other one to bjj are relatively clean and
   their rate will be large at the LHC ($\sim$0.1~Hz at low luminosity).
   These same event samples will also be used to measure the top mass. 
 
\item The knowledge of the absolute luminosity, which will
   contribute to the uncertainty on all cross section measurements.
Several methods are presently envisaged to determine 
   the absolute luminosity at the LHC~\cite{ATLAS,CMS},
   one of them being the measurement 
   of the rate of well-known processes such as
  W  and Z production. For the time being, the expected precision  
from the various methods  
is about 10\%. If this will be the case, then
  the luminosity uncertainty will be the dominant systematic error
  on most cross section measurements at the LHC. 
   Improved theoretical understanding of the 
  W  and Z production mechanism will therefore be extremely useful
  to reduce the luminosity uncertainty to a 
  more ambitious goal of  $\leq$~5\%.
    
\end{itemize}

\section{Measurement of the W mass\label{wmass}}
   
   At the time of the LHC start-up, the  W  mass
  will be known with a precision of $\sim$30~MeV from measurements
  at the TeVatron~\cite{Tev2000} and LEP2~\cite{LEP2}.    
    The motivation to improve this result is mainly that precise
   measurements of the W  mass, of the top mass and of the Higgs
   mass (SM Higgs boson or h boson in the MSSM) will provide 
   stringent tests of the consistency of the underlying theory. 
    With a top mass measured with an accuracy of  $\sim$1.5~GeV, as described
 in Sect.~\ref{top}, the W  mass should be known with a matching precision 
   of 15~MeV, in order not to become the dominant source of
  uncertainty in the test of the radiative corrections.
    Such a precision, which is
   beyond the sensitivity of LEP2 and TeVatron, should constrain 
   the Higgs mass to better than 30\%.
     
    At hadron colliders, the W  mass is obtained from the
   distribution of the W  transverse mass, that is the invariant mass of
   the W  decay products evaluated in the plane transverse to the beam. This
   is because the longitudinal component of the neutrino momentum
   cannot be measured in a pp or $\rm{p\overline{p}}$ collider. 
    On the other hand, the transverse momentum of the neutrino can be
   deduced from the transverse momentum imbalance in the calorimeters.
    The transverse mass distribution, and in particular its trailing edge,
   is sensitive to the value of the W  mass. 
    Therefore, by fitting the experimental
   distribution to Monte Carlo spectra obtained 
   for different values of $\MW$, it is possible to deduce
   the mass value which is preferred by the  data. 

     At the LHC, sixty million well-reconstructed W$\rightarrow\ell\nu$
   decays (where $\ell=\rm e$ or $\ell=\mu$) should be collected by each experiment 
   in one year of data taking at low luminosity (a statistics fifty times larger than
   that expected at the TeVatron Run II).      
     The statistical error on the W  mass measurement is therefore 
    expected to be small ($<$~2~MeV). 
     The systematic error will arise mainly from the Monte Carlo reliability
     in reproducing the data, i.e., the physics and the detector performance. 
     Uncertainties related to the physics are for instance the limited knowledge
     of the W $\pt$ spectrum, of the structure functions, 
     of the W  width and of the W radiative
     decays.  Uncertainties related to the detector
     are for instance the already-mentioned  absolute lepton scale and the 
     knowledge of the detector energy/momentum resolution and response to 
     the recoil. Many of these uncertainties will
     be constrained {\it in situ} by using the high-statistics
     sample of leptonic Z decays. This sample
     will be used for instance to set the lepton scale, 
     to determine the detector resolution and to model the detector response
     to the system recoiling against the W and the W $\pt$ spectrum. 
        
    Preliminary estimates of the expected uncertainties on the W  mass 
   measurement in ATLAS, based in part on extrapolating from TeVatron results,     
   are presented in Table~\ref{tab:wmass}.             
\begin{table*}[tbh]
\setlength{\tabcolsep}{1.5pc}
\newlength{\digitwidth} \settowidth{\digitwidth}{\rm 0}
\catcode`?=\active \def?{\kern\digitwidth}
\caption{Expected contributions to the systematic uncertainty
 on the W  mass measurement in ATLAS
 for each lepton family and for an integrated luminosity of
 10~fb$^{-1}$. The corresponding uncertainties 
 of the CDF measurement obtained in Run 1B~\cite{Wagner} 
 are also shown for comparison.\label{tab:wmass}}
\footnotesize
\begin{tabular*}{\textwidth}{@{}l@{\extracolsep{\fill}}lrrl}
\hline
         &               &               & \\[-.5pc]
Source                 & $\Delta m_W$ (CDF) & $\Delta m_W$ (ATLAS) &
         Comments \\
         &               &               & \\[-.5pc]
\hline
         &               &               & \\[-.5pc]
Lepton E/p scale       &  75 MeV      & 15 MeV      & TeVatron Run II: $<$ 40 MeV \\
Lepton E/p resolution  &  25 MeV      &  5 MeV      & Resolution known to $<\pm$1.5\% \\
Structure functions    &  15 MeV   &$<$ 10 MeV      & Constrained with LHC data \\
$\pt^{\rm W}$          &  20 MeV      &  5 MeV      & Constrained with
$\pt^{\rm Z}$ spectrum \\
Recoil model           &  33 MeV      &  5 MeV      & Constrained with Z data  \\
W  width               &  10 MeV      &  7 MeV      & $\Delta\GW=30$~MeV from TeVatron Run II\\
Radiative decays W$\to\ell\nu\gamma$       &  20 MeV   &$<$ 10 MeV
         & Better theoretical calculations   \\
         &               &               & \\[-.5pc]
\hline
         &               &               & \\[-.5pc]
Total                  & 113 MeV      &  $<$ 25 MeV & Per lepton
         species, per experiment   \\
         &               &               & \\[-.5pc]
\hline
\end{tabular*}
\end{table*}           
  As it is today at the TeVatron~\cite{CDF,D0}, also at the LHC the dominant 
 uncertainty will originate from the  calibration of the absolute lepton energy scale. 
  For the W  mass to be measured to better than 20~MeV,
   the lepton scale has to be known with a
 precision of 0.02\%, as already mentioned, which represents the most serious
 challenge for this measurement. It is interesting to note
that a very high precision on the lepton
 scale ($\sim$0.04\%) is also required
 at the TeVatron Run II 
 to reach the foreseen $\sim$40~MeV accuracy on $\MW$. The realization
of such a stringent requirement will represent a good benchmark
for the LHC experiments.
 
   All other systematic uncertainties are expected to be smaller
 than 10~MeV. Therefore, by combining both ATLAS and CMS
 and both channels (electrons and muons), it should be possible to obtain 
 a total error of 15~MeV in the initial, low luminosity phase. 
  Improved theoretical calculations, in particular
 of the W  $\pt$ spectrum and of the impact of 
 radiative corrections, will be needed to achieve this goal. 
 
Another possible method for measuring the W  mass
is based on the $\pt$ distribution of the charged lepton from the
W leptonic decay, that is
characterised by a Jacobian peak at $\pt^\ell\sim\MW/2$.
This method is weakly affected by pile-up and can thus
be used also in the high luminosity environement.
However, it is strongly dependent on the W $\pt$ spectrum
and therefore requires a very precise theoretical 
knowledge of the W $\pt$ distribution.
\section{Drell-Yan production of lepton pairs\label{Drell-Yan}}

\begin{figure}[tb]
\vspace{9pt}
\mbox{\epsfig{file=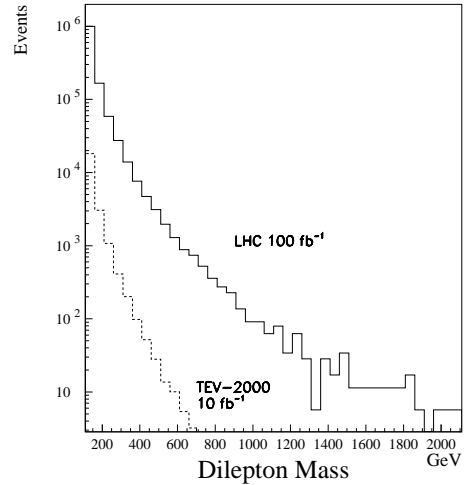,width=6.5cm}}
\caption{Expected number of Drell-Yan events for the TeVatron Run II
and LHC per
lepton channel and per experiment,
 as a function of the dilepton invariant mass.
} 
\label{fig1}      
\end{figure}

The production of lepton pairs via $s$-channel exchange of photons
or Z bosons is characterised by a very clean and distinctive
experimental signature, a pair of well isolated leptons with opposite
charge. At the LHC, the range of explored lepton invariant masses 
will be considerably extended, with respect to the presently
accessible region. This is shown in
Fig.~\ref{fig1}, which displays the expected number of events
per lepton channel after rapidity and $\pt$ cuts 
for one experiment at the TeVatron Run II and at the LHC,
as a function of the dilepton invariant mass.
Deviations in the expected behaviour can reveal new physics
(resonance formation, contact terms, etc.) or, if no signal of new physics
is visible, one can take advantage of the large statistics
to constrain parton densities and parton-parton luminosity functions~\cite{Dittmar}.
A prerequisite for the success of this program is that the
theoretical knowledge of the Drell-Yan process matches the
expected experimental accuracy.

\begin{figure}[tb]
\vspace{9pt}
\mbox{\epsfig{file=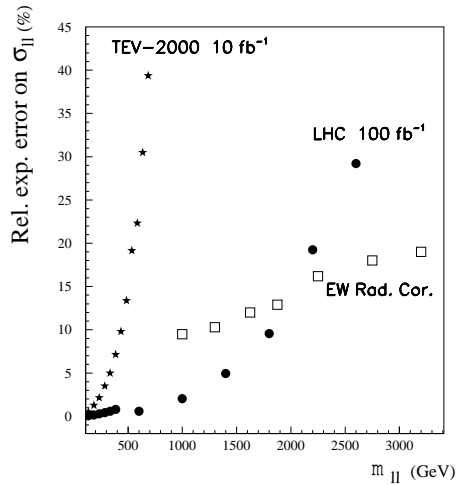,width=6.5cm}}
\caption{Relative statistical precision on the Drell-Yan cross
section (in \%) for an experiment at
the TeVatron Run II and at the LHC as a function of the dilepton
invariant mass.
Also shown is a calculation of the electroweak radiative corrections
 based on~\cite{DYrad}.
} 
\label{fig2}      
\end{figure}

The main observables of interest are the total cross section
and the forward-backward asymmetry which are both functions
of precisely measurable quantities, the invariant mass and rapidity
of the dilepton system.
Figure~\ref{fig2} shows the relative statistical precision on the cross section
measurement for one experiment at the TeVatron Run II and at the LHC
as a function of the dilepton invariant mass. 
Also shown for comparison is the result of
a complete one-loop parton
cross section calculation~\cite{LHCEW,DYrad} of the
electroweak radiative corrections after folding with the probability
density functions. The LHC will be able to probe such corrections
up to approximately 2~TeV. However, only the statistical error was
considered in this study. The uncertainty related to the luminosity
measurement will deteriorate the experimental precision by a few \%.
This example shows the importance of devoting the
 necessary effort, both on the theoretical and experimental side,
to achieve a precision of 5\% or better on the knowledge
of the absolute luminosity.

A precise determination of the effective electroweak mixing angle
$\sineff$ could be performed at the LHC
by measuring the forward-backward asymmetry A$_{\rm FB}$ in dilepton
production near the Z pole. The Z$\to\ell^+\ell^-$ cross section
is $\sim$1.5~nb for each lepton flavour, 
resulting in a very large number
of Z$\to\ell^+\ell^-$ events which, in principle, could be used to
measure $\sineff$ with a very small statistical error.
The latest combination of several asymmetry
measurements at LEP and SLD leads to an absolute uncertainty on $\sineff$ of
$1.7\times10^{-4}$~\cite{lepew}. Can systematic effects be controlled
with a comparable precision at the LHC?

The measurement of A$_{\rm FB}$ requires tagging
of the original quark direction which is unknown in pp collisions
and can only be extracted from the
kinematic properties of the dilepton system.
Events with a large rapidity of the lepton pair $y(\ell^+\ell^-$)
originate from collisions where at least one of the partons carries
a large fraction $x$ of the proton momentum. Since valence quarks
carry on average a higher momentum than sea antiquarks,   A$_{\rm FB}$
is signed according to the sign of  $y(\ell^+\ell^-$).
Recent preliminary studies~\cite{LHCEW} indicate that a statistical precision
on $\sineff$ better than $10^{-4}$ may be obtained 
by extending the detection of one of the two leptons
(in the electron channel)
to the rapidity range $2.5<|y_{\ell}|<4.9$ 
because  A$_{\rm FB}$
is significantly larger in this region. 
Such a strategy is based on a moderate e/$\pi$ separation capability
($\pi$ rejection of 10--100) in the
forward calorimeter, which however still needs to be proven.
The main systematic effect on the measurement of
$\sineff$ originates from the uncertainty on the parton distribution
functions which affects the lepton acceptance as well as the results
of the radiative correction calculations.
It is far from obvious that this uncertainty can be
brought down to the desired level of precision. However,
new measurements from HERA, TeVatron and, ultimately, from the
LHC itself, will certainly improve the uncertainty on the parton
distribution functions and may render this measurement possible.

\section{Measurement of the Triple-Gauge Couplings\label{TGC}}
  
  The study of Triple-Gauge Couplings (TGCs), that is couplings of the type
$\rm{WW}\gamma$ or WWZ, provides a direct test of the non-abelian structure of the 
SM gauge group and at the same time may yield hints for new physics, since
many new processes are expected to give anomalous contributions to the
triple-gauge vertices. 
  This sector of the Standard Model is often described by five
parameters: $g_1^{\rm Z}$,   $\kappa_{\rm Z}$, $\kappa_{\gamma}$,
$\lambda_{\rm Z}$, $\lambda_{\gamma}$.
New physics could show up as deviations of these parameters 
from their SM values (zero for the $\lambda$ parameters and
one for the $\kappa$ and $g$ parameters). 
The LHC has a large 
potential for testing the TGCs because the sensitivity to  
the anomalous contributions is enhanced at high centre-of-mass
energies, particularly for $\lambda$-type TGCs.

Triple-Gauge couplings  will give rise to 
gauge boson pair production, e.g., W$\gamma$, WZ and WW production.
The first two processes are characterised by 
relatively clean final
states, containing one lepton and one photon or
three leptons, respectively. The third process is less promising, since it
suffers from the large $\rm{t\overline{t}}$ background.
    
\begin{figure}[tb]
\vspace{-6.5pt}
\mbox{\epsfig{file=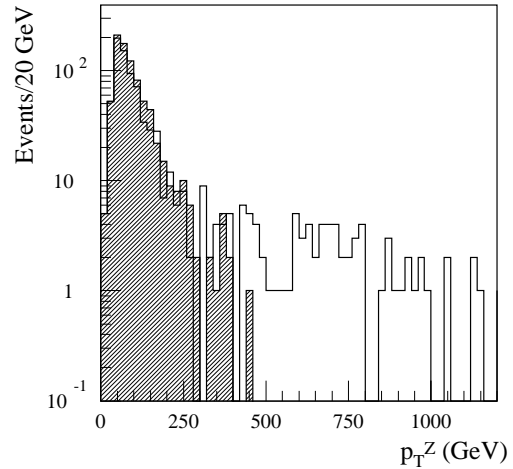,height=7.2cm}}
\caption{Reconstructed 
$\pt$ of the Z boson in WZ events, as 
expected in ATLAS
for an integrated luminosity of 30~fb$^{-1}$. 
The shaded and white histograms show, respectively, the
SM expectation and the distribution obtained for $\Delta g_1^{\rm Z}=0.05$.  
} 
\label{fig:tgc}      
\end{figure}

Anomalous TGCs can affect both the total cross section and the
shape of the differential distributions. This is illustrated in
Fig.~\ref{fig:tgc} that shows, as an example, the reconstructed $\pt$ spectra of
the Z boson in WZ events for the SM and in presence of non-standard
couplings. An excess of events in the high-$\pt$ tail is clearly visible
in the case of anomalous couplings.

The expected statistical precision at 95\%~CL from
single parameter fits to the total cross section and to distributions such as 
the one of Fig.~\ref{fig:tgc} is summarised 
in Table~\ref{tab:tgc}, as obtained by ATLAS assuming an integrated 
luminosity of 30~fb$^{-1}$. 
\begin{table}[tb]
\caption{
Sensitivity limits (95\%CL) from single
parameter fits to a given coupling for one LHC experiment assuming an integrated
luminosity of 30~fb$^{-1}$. A form factor $\Lambda$=10~TeV has been assumed
in deriving these results.\label{tab:tgc}}
\begin{center}
\footnotesize
\begin{tabular}{cl}
\hline
         &             \\[-.4pc]
  Coupling  & 95\%CL   \\
         &             \\[-.4pc]
\hline
         &             \\[-.4pc]
$\Delta\kappa_{\gamma}$ & 0.035\\
$\lambda_{\gamma}$      & 0.0025\\
$\Delta g_1^{\rm Z}$    & 0.0078\\
$\Delta\kappa_{\rm Z}$  & 0.069\\
$\lambda_{\rm Z}$       & 0.0058\\
         &             \\[-.4pc]
\hline
\end{tabular}
\end{center}
\end{table} 
These results based on a modest total integrated luminosity
already improve on the 
final TeVatron and LEP2 precision expected to be in the range between $\sim$0.1 and $\sim$0.01
depending on the couplings. Systematic uncertainties due, for example,
to higher order QCD corrections, structure functions, etc., are
currently under study but are expected to be small. 
%

\section{Precision measurements in the top quark sector\label{top}}

   Because of its large mass and width and of the 
   special r\^ole it plays in  radiative corrections, 
   the top quark is a very peculiar fermion. Precision
   measurements in the top sector are therefore important 
   to get more clues on the origin of the fermion mass hierarchy. 
  
   At the LHC, top quark measurements 
  will benefit from very large statistics, 
    so that not only the mass and the production cross section,
  but also branching ratios, couplings and exotic decays will be studied
  in detail. It should also be noticed that   
  $\rm{t\overline{t}}$ production is expected to be the main background to  
  new physics processes, such as several possible channels arising
  from the
  production and decay of  Higgs bosons and of supersymmetric particles. 
   Furthermore, top events will
  be used to calibrate the calorimeter jet scale, as already mentioned 
  in Sect.~\ref{key}. 
 
   The $\rm{t\overline{t}}$ production cross section is expected to be 
  $\sim$800~pb at the LHC, to be compared with $\sim$7~pb at the TeVatron. 
  Taking into account also the higher luminosity, the LHC should be able
  to collect, in the initial phase at low luminosity, an event sample
  at least 1000 times larger than the one expected in the future
  at the TeVatron.  

   In the year 2005, the top mass should be known with a
   precision of 3~GeV or better from measurements at the
   TeVatron~\cite{Tev2000}. At the LHC 
   the best channel for the top mass measurement 
   will most likely be $\rm{t\overline{t}}$ production
   with one W  decaying leptonically and the other  
   one hadronically. The top mass will be determined from the hadronic
part of the decay, as the invariant mass of the three jets originating
from the same top ($\Mt=m_{\rm jjb}$). The leptonic top decay will be used to
tag the event by exploiting the high $\pt$ lepton and large $\Et^{\rm miss}$.

The statistical error 
is expected to be negligible (smaller than 100 MeV), 
therefore the precision will be limited by the systematic error.
The 1\% uncertainty on the absolute jet scale should translate
into an uncertainty  smaller than 1~GeV on the top mass. 
The effect of final-state 
gluon radiation is estimated to lead to an uncertainty of $\sim$1~GeV.
Other sources of systematic uncertainties 
(such as, for example, those related to b-fragmentation, 
initial state radiation, background, etc.) are expected to be smaller.
  
All together, a total uncertainty smaller than 1\% should be achieved. 
This precision may be further improved by using $\rm{t\overline{t}}$ pairs
produced with very high $\pt$. In this case, the two top quark decay products
are well separated in two opposite hemispheres, so that
the mass measurement should be less sensitive to the details
of the jet reconstruction method,
to the choice of the fragmentation model and to the combinatorial
background from gluon radiation. 

\begin{figure}[tb]
\mbox{\epsfig{file=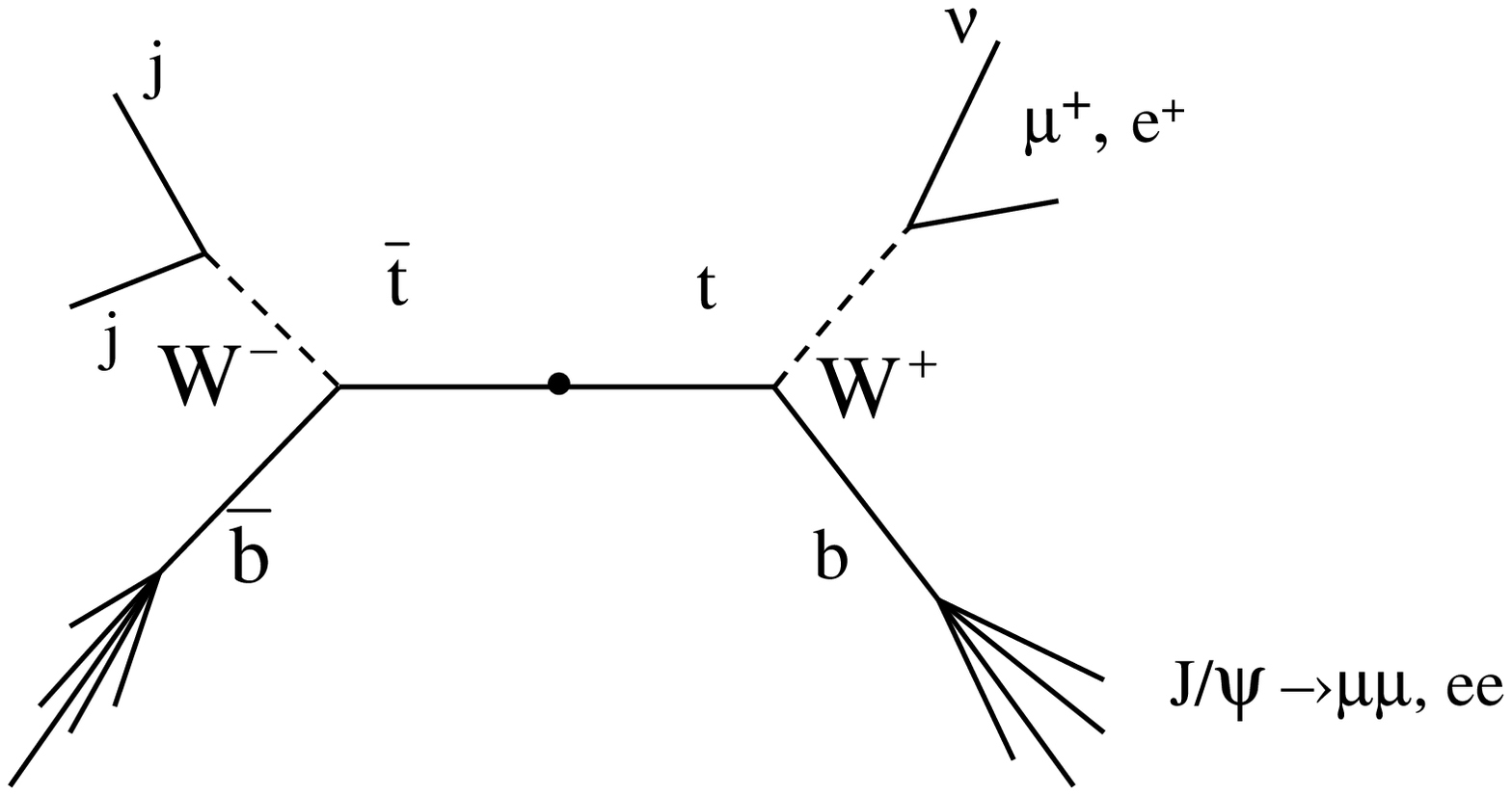,width=6.5cm}}
\caption{
Diagram for top decaying into a J/$\Psi$+lepton final state.
} 
\label{fig:tdecay}      
\end{figure}

Another interesting idea for measuring $\Mt$ proposed by
CMS~\cite{Jpsi,Khar} is based on the decay 
${\rm t\to J/\Psi+X}$ displayed in  Fig.~\ref{fig:tdecay}.
In this case one takes advantage of the correlation between $\Mt$
and the invariant mass of the  $\rm J/\Psi$ and lepton originating from
the same top, that is
enhanced by the presence of a heavy object, the  $\rm J/\Psi$, 
carrying a large fraction of the b momentum.
The small branching
fraction characterising this channel, of O(10$^{-5}$), is compensated
by the clean final state which can be exploited also
at the highest luminosities. The main systematic limitation of such
a measurement originates from the uncertainty on the fragmentation
function of the B hadrons contained in the b jet.
Current preliminary studies suggest that such an analysis
might lead to an error on $\Mt$ comparable or smaller
than the one obtained from the single lepton plus jets channel.
  
In any case, the possibility of selecting different samples, 
each characterised by different systematic uncertainties,
will allow useful cross-checks of the top mass determination.

  Examples of other measurements which can be performed in the top sector
  are:
 
\begin{itemize}

\item The $\rm{t\overline{t}}$ production cross section 
should be determined with a precision  $\leq 10\%$, dominated by the uncertainty on the absolute luminosity. 
 
\item Single top production via the weak interaction,
a process not yet observed,
should allow 
the CKM matrix element $V_{\rm tb}$  to be measured with a precision of $10\%$ or better.

\item  Upper limits at the level
of $10^{-4}-10^{-5}$ on the FCNC
couplings tVc and tVu with V=Z, $\gamma$, g  
should be set with 100 fb$^{-1}$, improving by a factor of
at least 10 the TeVatron sensitivity. 
   
%
\item  A sensitivity of $\sim$3\% on the branching ratio
${\rm BR(t\to \rm{bH}^{\pm})}$ should be reached,
by searching for an excess of $\tau$ production in $\rm{t\overline{t}}$ events,
due to the decay $\rm t\to\rm{bH}^{\pm}$,
followed by ${\rm H}^{\pm}\to \tau\nu$.
This would allow $\rm H^{\pm}$ masses below $\Mt-20$~GeV to be probed for most of
the tan$\beta$ range.
\end{itemize}

More details can be found elsewhere~\cite{TDR,LHCtop}.  

\section{Precision measurements in the Higgs sector\label{Higgs}}

  The LHC discovery potential for a SM Higgs boson is well known since 
  a long time~\cite{ATLAS,CMS}. 
  After less than two years of data taking at low luminosity,
 a signal significance over the background larger than 5~$\sigma$ 
 is expected over the mass range between 115~GeV (approximate LEP2 lower bound)
and 1~TeV (upper bound predicted by theory) by combining both experiments.  

   Assuming that a SM Higgs boson 
  will be found at the LHC,  the question of the precision
  with which the ATLAS and CMS experiments will be able
  to measure the Higgs parameters 
  (e.g. mass, width, cross section, couplings) can be addressed.   

\begin{figure}[tb]
\centering
\mbox{\epsfig{file=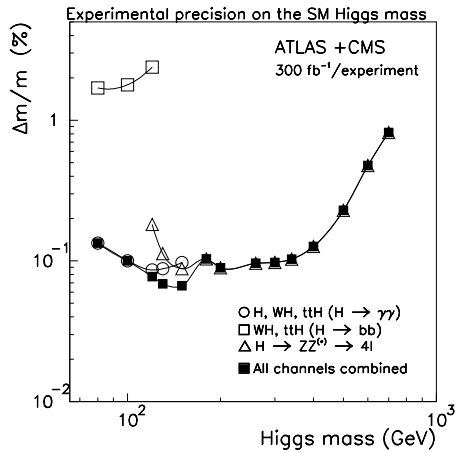,width=6.5cm}}
\caption{Expected fractional errors on the measured  Higgs mass 
 at the LHC, as a function of $\MH$. 
  The different symbols 
 indicate different production and decay channels.} 
\label{fig:Hmass}      
\end{figure}

    Figure~\ref{fig:Hmass} shows the expected uncertainty
   on the measurement of the Higgs mass, as obtained by combining 
   both experiments and for an integrated luminosity of 
   300~fb$^{-1}$ per experiment. 
    All expected experimental uncertainties are included in these results,
   i.e., the statistical error and the systematic error due to  
   the uncertainty  on the absolute energy scale and on the 
   background subtraction. It can be seen that 
   a precision of 0.1\% can be obtained up to
   $\MH\simeq 500$~GeV. For larger masses the precision
   deteriorates because the statistical error increases. 
   Theoretical uncertainties are not taken into account in these results.
   Whereas the uncertainty arising  
   from the knowledge of the structure functions is expected
   to be small, other effects may have a non-negligible impact. 
   For instance, for large Higgs masses, when the Higgs width becomes broad, 
   interference effects between the resonant and the non-resonant 
   cross section are expected
   to produce a downward shift of the Higgs mass peak~\cite{baur}. 
   Accurate input from theory will therefore be needed in this case. 

\begin{figure}[tb]
\centering
\mbox{\epsfig{file=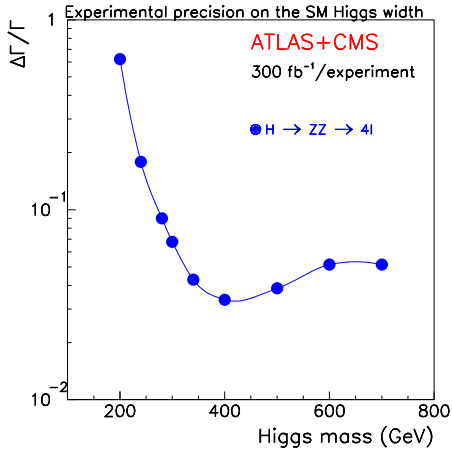,width=6.5cm}}
\caption{Expected fractional errors on the measured  Higgs width 
 at the LHC, as a function of $\MH$.} 
\label{fig:Hwidth}      
\end{figure}

   The Higgs width can be directly extracted  
from a measurement of the  width of the reconstructed Higgs peak
after unfolding of the detector resolution. This, however,
   will only be possible
  for Higgs masses larger than about 200~GeV, above which
  the intrinsic width of the resonance is comparable to or larger than 
  the expected experimental resolution. For smaller masses the Higgs width is 
  too narrow to be measured with the direct method.
   Figure~\ref{fig:Hwidth} shows the 
  expected uncertainty on the Higgs width, as a function of
  the Higgs mass, as obtained by combining both experiments and for
  an integrated luminosity of 300 fb$^{-1}$ per experiment. 
The precision on $\GH$ improves up to Higgs masses of approximately 300~GeV, after
which the total resolution is dominated by the intrinsic width.
In the mass range 300--700~GeV an approximately
constant precision of $\sim$5\% is obtained.
This is the region where the best discovery channel is
  $\rm {H\rightarrow ZZ}\rightarrow 4\ell$. 
   The measurement of the Higgs width to such an accuracy requires a very
  good knowledge of the detector energy and momentum resolution.
  The detector resolution is expected to be determined with a 
  precision better than 1.5\% from the measurement of the Z width.
   This error, which is dominated by the systematic uncertainty on the
  radiative decays of the Z, has been 
  included in the results shown in Fig.~\ref{fig:Hwidth}.  
    
    A measurement of the Higgs production rate in a given channel
   provides a measurement of the production cross section multiplied
   by the branching ratio in that channel $\rm \sigma\cdot BR$. 
    At the LHC the precision will be mainly limited by 
    the uncertainty on the
    luminosity. For a luminosity uncertainty of 5\%, 
    $\rm \sigma\cdot BR$ should be measured with a typical accuracy of about 
   7\% over the mass region 100$<\MH<$700~GeV, by combining
   both experiments and with an integrated luminosity 
   of 300~fb$^{-1}$ per experiment. This accuracy would degrade
   to $\sim$12\% for a luminosity uncertainty of 10\%.
  
\begin{figure}[bt]
\vspace{0.5cm}
\centering
\mbox{\epsfig{file=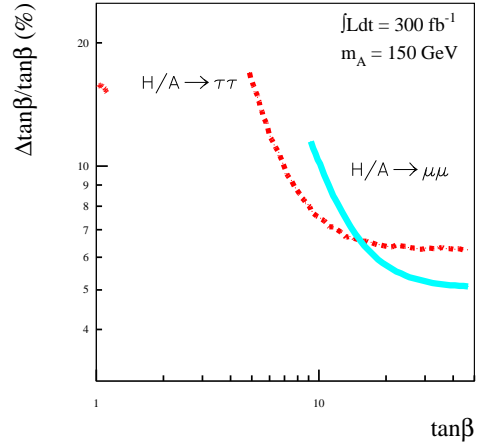,width=6.5cm}}
\caption{Expected fractional errors on tan$\beta$
in ATLAS assuming $\MA=150$~GeV and a total integrated
luminosity of 300~pb$^{-1}$.} 
\label{fig:tanbeta}      
\end{figure}

The rates of the heavy Higgs bosons of the MSSM (H/A) provide 
good sensitivity to tan$\beta$~\cite{TDR}. As shown in Fig.~\ref{fig:tanbeta},
from the measurements of the rate of H/A~$\to\tau\tau$,
tan$\beta$ can be determined with a $\pm15\%$~($\pm6\%$) precision for
tan$\beta$=5~(40), assuming $\MA=150$~GeV, a total integrated
luminosity of 300~pb$^{-1}$ and one experiment. A somewhat better precision is obtained
at higher tan$\beta$ values by analysing the $\mu\mu$ final state.

    From the measured $\rm \sigma\cdot BR$, one can deduce the Higgs 
   branching ratio in a given channel if the Higgs production cross section
   is known from theory. Without theoretical assumptions, it is
   still possible to measure ratios of branching ratios, and therefore
   ratios of Higgs couplings to fermions and bosons. 
Work in this sector has just started, therefore only   
a few examples are given below.  It is assumed that 
   both experiments are combined and that the integrated
   luminosity is 300~fb$^{-1}$ per experiment.

\begin{itemize}
\item By looking at the associated Higgs production 
 ($\rm {t\overline{t}H}+\rm{WH}$) and by measuring 
 the rate of events where the Higgs decays
 to $\gamma\gamma$ divided by the rate of 
 events where the Higgs decays to $\rm{b\overline{b}}$, it is
 possible to obtain the ratio between the 
 branching ratio to $\gamma\gamma$  and the branching ratio
 to $\rm{b\overline{b}}$. This measurement can be performed
 over the mass range 80$\leq\MH\leq$120~GeV and the expected
 precision is about 15\%. 
\item By comparing the rate of Higgs bosons produced in association
 with a $\rm{t\overline{t}}$ pair to the rate of Higgs bosons produced in
 association with a W  (with the Higgs decaying to $\gamma\gamma$
 or to $\rm{b\overline{b}}$), it is possible to measure the ratio
 of couplings $\rm{(t\overline{t}H/WWH)^2}$. This measurement
 can be performed over the mass range
 80$\leq~\MH~\leq$120~GeV and the expected precision is about 15\%.  
\item By measuring the ratio between the ${\rm H}\rightarrow\gamma\gamma$ rate
 and the ${\rm H}\rightarrow 4\ell$ rate, it is possible
 to obtain the ratio between the
 branching ratio to $\gamma\gamma$ and the
 branching ratio to $\rm{ZZ}^*$. This measurement can be performed
 over the mass range $120\leq \MH\leq150$~GeV and the expected
 precision is about 7\%. 
\end{itemize}
In all the above cases, the error is dominated by the statistical
uncertainty, since the systematic uncertainty on the absolute luminosity 
cancels out when ratios of rates are considered.

\section{Precision measurements in the SUSY sector\label{SUSY}}

 It is well known that, if SUSY exists close to the electroweak
 scale, it will be easy to discover at the LHC for 
  $\tilde{\rm q}$ and $\tilde{\rm g}$
  masses up to about 3~TeV and 
  almost irrespectively of the model parameters. 

  Assuming that SUSY will be discovered at the LHC, will
 the ATLAS and CMS experiments be able to perform
 precision measurements in SUSY final states, i.e., determine the
 particle masses and their couplings, and therefore extract the fundamental 
 parameters of the theory ?
  The answer to this question is a priori not obvious:   
  in R-parity conserving scenarios all SUSY events
  contain in the final state the
   two lightest neutralinos, which are stable and
  weakly interacting and hence escape detection. 
   Therefore, in general there are not enough kinematic constraints 
   to reconstruct mass peaks.  
   In order to investigate this issue, 
  five points in the parameter 
  space of Minimal SuperGravity (SUGRA~\cite{SUGRA}) were studied. 
  
   SUGRA is a model with only five parameters:
   a common scalar mass at the GUT scale ($m_0$), a common gaugino mass
   at the GUT scale ($m_{1/2}$), the ratio of the vacuum
   expectation values of the two Higgs doublets (tan$\beta$), a common 
   trilinear term  at the GUT scale ($A_0$) and the sign of the Higgsino
   mass parameter ($\mu$).    

    The strategy adopted by ATLAS in the study of the five SUGRA points
   is the following. First, find an inclusive SUSY signal over the SM
   background. Second, try to 
   isolate exclusive, therefore clean,
    channels where masses or combination of masses can be measured
   from kinematic distributions;
this is possible  because in most cases the expected event samples
   are large. Finally, perform a global fit of the model to all 
   experimental measurements
   and extract the fundamental parameters of the theory, very much
   in the same way as the LEP experiments have done to test the SM
   predictions and to determine indirectly the top, W and Higgs masses.

    As an illustration of the method and of the results
   which can be achieved, one of the five points (``Point~5")
   is discussed here in some detail.   
   Point~5 is characterised by the following values of the SUGRA fundamental 
  parameters: $m_0$=~100~GeV, $m_{1/2}=$~300~GeV, $A_0$~=~300~GeV, 
  tan$\beta$~=~2.1, sign$\mu$~=~+. 
   The masses of some of the corresponding 
   SUSY particles are listed in 
   Table~\ref{tab:spart}. 
      
\begin{table}[tb]
\caption{Masses of some representative SUSY particles 
 in Point~5.\label{tab:spart}}
\begin{center}
\footnotesize
\begin{tabular}{lc}
\hline
Particle & Mass (GeV)  \\
\hline
$\tilde{\rm g}$        & 770 \\
$\tilde{\rm q}_L$      & 690 \\
$\tilde{\rm q}_R$      & 660 \\
$\tilde{\rm t}_1$      & 490 \\
$\tilde{\ell}_L$   & 240 \\
$\tilde{\ell}_R$   & 157 \\ 
$\chi^0_1$         & 121 \\
$\chi^0_2$         & 232 \\
$\rm h$            & 93  \\
$\rm H$            & 640 \\
\hline
\end{tabular}
\end{center}
\end{table}

   The total cross section, which is mainly determined by the 
  $\tilde{\rm q}$ and $\tilde{\rm g}$ masses since 
  $\rm {\tilde{q}\tilde{q},\ \tilde{q}\tilde{g}}$ and  $\rm \tilde{g}\tilde{g}$ 
  production dominates, is about 20~pb. The second lightest neutralino
  decays to the lightest Higgs boson and the lighest neutralino
   ($\chi^0_2\rightarrow {\rm h}\chi^0_1$) with a branching ratio of almost
   70\%. It can also decay to slepton-lepton pairs  
   ($\chi^0_2\rightarrow\tilde{\ell}_R\ell$) with a branching
   ratio of about 10\% per lepton species, since sleptons
   are relatively light for this choice of the SUGRA parameters.  

%

\begin{figure}[bt]
\centering
\mbox{\epsfig{file=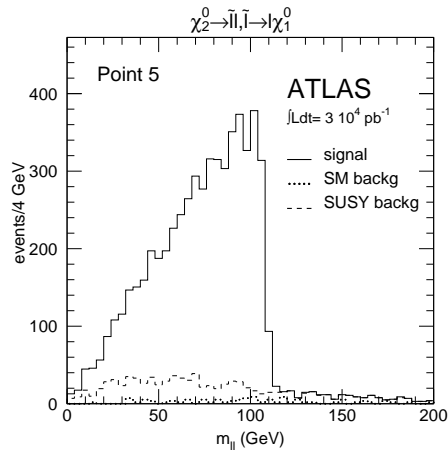,width=6.5cm}}
\caption{Invariant mass distribution of lepton pairs
 in the final state for SUSY events at Point~5 selected as described
 in the text, as expected in ATLAS after three years of data taking
 at low luminosity.} 
\label{fig:slep}      
\end{figure}

  As an example, the production of 
 $\rm \tilde{q}_L\tilde{q}_R$, followed by the 
 decays ${\rm \tilde{q}_L\to q}\chi^0_2$, 
$\chi^0_2\rightarrow\tilde{\ell}_R\ell$, 
 $\tilde{\ell}_R\rightarrow\ell\chi^0_1$, 
can be selected in an inclusive way  by requiring two 
  leptons in the final state with the same flavour and opposite
  sign, large $E_{\rm T}^{\rm miss}$ and
  jet multiplicity (the last two cuts are needed to reject the SM background). 
   The resulting invariant mass distribution of the two leptons in the final state
  is shown in Fig.~\ref{fig:slep}. A clear signal is visible above the background. 
   The mass distribution shows a very sharp end-point, which is due 
   to the kinematic properties of the decay and depends
   on the masses of the involved particles 
   (the two lightest neutralinos and the slepton) through
   a simple kinematic relation. The position of the end-point can be measured with
   a precision of 500~MeV (0.5\%) with an integrated luminosity of 30~fb$^{-1}$, 
    thus providing a combined constraint on the three masses mentioned above.  

 More details and further examples can be found 
 elsewhere~\cite{TDR,giacomo}. 
 

 Table~\ref{tab:point5} summarises the various measurements of particle
 masses which can be performed in Point~5, together with the expected
 precisions in ATLAS for two integrated luminosities.
 In all cases, the ultimate precision is 
 in the range between a few percent and a few permil. 
  Furthermore, many SUSY particles (h, $\rm \tilde{q}_L$,
 $\rm \tilde{q}_R$, $\rm \tilde{g}$, $\rm \tilde{t}_1$, $\rm \tilde{b}_R$, 
 $\chi^0_2$, $\tilde{\ell}_{\rm L}$, $\tilde{\ell}_{\rm R}$) will be
  directly observable at the LHC. 

%
\begin{table*}[tb] 
\setlength{\tabcolsep}{1.5pc}
\settowidth{\digitwidth}{\rm 0}
\catcode`?=\active \def?{\kern\digitwidth}
\caption{Expected uncertainties on the measurements of SUSY particle
 masses at Point~5 in ATLAS for two different  
 integrated luminosities.\label{tab:point5}}
\footnotesize
\begin{tabular*}{\textwidth}{@{}l@{\extracolsep{\fill}}lccc}
\hline
Measurement & Expected value  & Error (\%)  & Error (\%) \\
            &   (GeV)         & 30~fb$^{-1}$  & 300~fb$^{-1}$  \\
\hline

$m_{\rm h}$                & 93         & $\pm$1.0  & $\pm$0.2 \\
$m_{\ell^+\ell^-}$ end-point      & 109 & $\pm$0.5  & $\pm$0.2 \\
$m_{\tilde{\ell}_{\rm R}}$   & 157      & $\pm$1.2  & $\pm$0.3 \\
$m_{\tilde{\ell}_{\rm L}}$   & 240      & $\pm$4    & $\pm$1 \\
$m_{\rm \tilde{q}_L}$      & 690        & $\pm$1.7  & $\pm$1 \\
$m_{\rm \tilde{q}_R}$      & 660        & $\pm$3    & $\pm$1.5 \\
$m_{\rm \tilde{g}}$        & 770        & $\pm$2.6  & $\pm$1.5 \\
$m_{\rm \tilde{t}_1}$      & 490        &           & $\pm$10 \\
\hline
\end{tabular*}
\end{table*}    
  
  The above experimental measurements can then be used to constrain the
 model and  its parameters. 
  The results of the global fit in the
 case of Point~5 are presented in Table~\ref{tab:fit}.  
\begin{table}[tb]
\caption{Expected uncertainty on the measurement of the fundamental
 SUGRA parameters at Point~5 in ATLAS for two different 
 integrated luminosities.\label{tab:fit}}
\begin{center}
\footnotesize
\begin{tabular}{lcc}
\hline
SUGRA parameter &  Error for        & Error for  \\
                &  30~fb$^{-1}$ & 300~fb$^{-1}$ \\
\hline
$m_0=100$~GeV      & $\pm$5~GeV  & $\pm 3$~GeV \\
$m_{1/2}=300$~GeV  & $\pm$8~GeV  & $\pm 4$~GeV \\
tan$\beta$=~2.1     & $\pm$0.11   & $\pm 0.02$  \\
\hline
\end{tabular}
\end{center}
\end{table}        
  The parameters
 $m_0, m_{1/2}$ and tan$\beta$ will be determined by ATLAS
  with a precision of a few percent after only three years 
 of data taking at low luminosity.  The sign of $\mu$ will also be
 unambiguously determined. The $A_0$ parameter will
 remain most likely unconstrained, because
 it has a very little influence on the phenomenology at the electroweak
 scale. Similar results were obtained  for the four other SUGRA 
 points~\cite{TDR}. 

   In conclusion, precision SUSY measurements will be possible at the LHC:
   many SUSY
   particles should be discovered, and many of their masses should be
   measured with precisions between a few permil and a few percent.
   Such a potential arises
   mainly from the large SUSY cross section and  the variety of
   signatures which are produced by the cascade decays of $\rm\tilde{q}$
   and $\rm\tilde{g}$.  The fundamental
   parameters of minimal SUGRA should be measured with precisions of 
   the order of 1\%.
   
    More generally, whatever the correct theory will be,
    the LHC experiments will be able to perform many model-independent
   observations and measurements, such as observations of 
   excesses of events with top quarks, b quarks, Z bosons,
   observation of $\rm h\to b\overline{b}$ peaks, 
    measurements of end-points and shapes of several types of mass spectra. 
     This will provide a large number  of experimental results,
    which should constrain quite general SUSY models, models with R-parity
    breaking, and Gauge-mediated SUSY-breaking theories~\cite{TDR}.

\section{Conclusions}

   In addition to its huge potential for the discovery of
  new physics, the LHC will allow a wealth of precision measurements
   to be performed in many sectors: 
  W/Z physics, Triple-Gauge Couplings,
  top physics, B physics, Higgs, Supersymmetry, etc. A few non-exhaustive
  examples have been presented. 
  
  In many cases, significant improvements on the future TeVatron and
  LEP results are expected after only one or two years of
  operation. 
  
   The statistical error will be negligible for most measurements, 
   and the precision will be limited by systematic effects.  
   Therefore, stringent requirements have been set on the
    design of the ATLAS and CMS experiments, 
    and on their performance in terms of energy and
   momentum resolution, response uniformity, 
   particle identification capability, etc.    
     If the experiments will behave as expected, the
   precision of many measurements will be limited by the  
   knowledge of the physics and not 
   by the detector performance. Therefore,
   improved theoretical calculations
   will be necessary to match the expected experimental
   accuracies.  


\end{document}